# Ellipsoidal anisotropy in elasticity for rocks and rock masses

Elasticité à anisotropie ellipsoïdale pour les roches et les massifs rocheux

A. Pouya
*Laboratoire Central des Ponts et Chaussées, 58 Bd Lefebvre, 75732, Paris Cedex 15, France*

M. Chalhoub
*Université Saint Esprit de Kaslik, Jounieh, Lebanon*

**RESUME:** Les modèles d'anisotropie ellipsoïdale en élasticité linéaire présentent plusieurs intérêts. D'abord, les solutions analytiques de beaucoup de problèmes classiques d'élasticité linéaire connues pour le cas de comportement isotrope peuvent s'étendre à certains types d'anisotropie ellipsoïdale à l'aide d'une transformation linéaire simple (Pouya 2000, Pouya & Zaoui 2006). Ces modèles permettent aussi une analyse simple des données expérimentales. Dans ce travail, nous montrons que ces modèles peuvent ajuster avec une bonne approximation le comportement de certaines variétés de roches et de massifs rocheux anisotropes.

**ABSTRACT:** One of the interesting features with the ellipsoidal models of anisotropy presented in this paper is their acceptance of analytical solutions for some of the basic elasticity problems. It was shown by Pouya (2000) and Pouya and Zaoui (2006) that many closed-form solutions for basic problems involving linear isotropic materials could be extended by *linear transformation* to cover a variety of "ellipsoidal" materials. This paper will describe two main varieties of ellipsoidal elastic models and show how well they fit the *in situ* data for sedimentary rocks; numerical homogenization results for several varieties of fractured rock masses will also be provided.

## 1. INTRODUCTION

In some anisotropic elasticity problems, information is available on the values of an elastic parameter in the various directions, to be used in identifying the elastic tensor. A typical case in Rock Mechanics involves deducing Young's modulus from simple compression tests in different directions or measuring the acoustic velocity in different directions on a sample (Homand *et al*. 1993; Francois *et al*. 1998). Based on the notion that material isotropy corresponds geometrically to the image of a sphere, an expression for anisotropy can naturally be sought through an ellipsoidal variation of a number of parameters in different directions. The uncertainty then lies in how to deduce the anisotropic tensor from this assumption. Saint Venant (1863) studied this specific question intensively by introducing the approximation of ellipsoidal indicator surfaces. The indicator surface of an elastic parameter $c$ is the polar diagram $c(\boldsymbol{n})$, where $\boldsymbol{n}$ is a unit vector and $c(\boldsymbol{n})$ the value of parameter $c$ in the material direction $\boldsymbol{n}$. In recent years, the concept of ellipsoidal anisotropy has been adopted as a guideline for the phenomenological modelling of *geomaterials* such as soils, rocks and concrete (Peres Rodrigues, 1970; Daley and Hron, 1979). Yet, the concept of anisotropic elasticity has at times been employed erroneously. For instance, Peres Rodrigues (1970) attempted using ellipsoids to fit, for several types of rocks, Young's modulus values measured along different directions. It has been shown (Pouya, 2007a) however that the Young's modulus indicator surface, i.e. the polar diagram of $E(\boldsymbol{n})$, can never be an ellipsoid (different from a sphere), hence the parameters fitted by this author do not define any possible elasticity tensor. The correct approach calls for fitting the diagram of $\sqrt[4]{E(\boldsymbol{n})}$ by an ellipsoid; this was performed by Saint Venant in 1863.

## 2. ELLIPSOIDAL MODELS

Let's define the elastic coefficient and Young's modulus in the $\boldsymbol{n}$ direction, where $\boldsymbol{n}$ is a unit vector defining direction in the material, respectively by:

$$c(\boldsymbol{n}) = (\boldsymbol{n} \otimes \boldsymbol{n}) : \mathbb{C} : (\boldsymbol{n} \otimes \boldsymbol{n})$$

$$E(\boldsymbol{n}) = [(\boldsymbol{n} \otimes \boldsymbol{n}) : \mathbb{S} : (\boldsymbol{n} \otimes \boldsymbol{n})]^{-1} \quad (1)$$

with $\mathbb{C}$ being the fourth-order elasticity tensor and $\mathbb{S} = \mathbb{C}^{-1}$. The main family of ellipsoidal materials considered herein has been defined by the condition that the indicator surface of $\sqrt[4]{c(\boldsymbol{n})}$, i.e. the surface given by polar equation $r(\boldsymbol{n}) = \sqrt[4]{c(\boldsymbol{n})}$, is ellipsoidal. This condition specifies a family of materials that depends on 12 independent parameters (Pouya, 2007a). The intersection of this family and the family of orthotropic materials in turn defines another family, denoted here by $\Phi_4$, that depends on 6 intrinsic parameters ($c_{11}$, $c_{22}$, $c_{33}$, $c_{12}$, $c_{13}$ and $c_{23}$) in the orthotropic coordinate axes, as described by the following relations between elastic coefficients:

$$\Phi_4 : \begin{cases} c_{44} = \dfrac{\sqrt{c_{22}c_{33}} - c_{23}}{2}, \quad c_{55} = \dfrac{\sqrt{c_{11}c_{33}} - c_{13}}{2}, \\ c_{66} = \dfrac{\sqrt{c_{11}c_{22}} - c_{12}}{2} \end{cases} \quad (2)$$

The second family to be taken into consideration in this study is the subfamily of $\Phi_4$, for which the indicator surface of $\sqrt[4]{E(n)}$ is also ellipsoidal. This family, denoted $\Psi$, is dependent upon four independent parameters $c_{11}$, $c_{22}$, $c_{33}$ and $\eta$; it is described by the following conditions:

$$c_{12} = \eta \sqrt{c_{11} c_{22}} \ , \quad c_{44} = \frac{1-\eta}{2} \sqrt{c_{22} c_{33}} \quad (3)$$

with other coefficients obtained by means of index permutation {1,2,3} and, correspondingly, {4,5,6}. This family can be equivalently defined by the four parameters $E_1$, $E_2$, $E_3$ and $\nu$, as follows, with $\Psi$:

$$\begin{bmatrix} \frac{1}{E_1} & \frac{-\nu}{\sqrt{E_1 E_2}} & \frac{-\nu}{\sqrt{E_1 E_3}} & & & \\ \frac{-\nu}{\sqrt{E_1 E_2}} & \frac{1}{E_2} & \frac{-\nu}{\sqrt{E_2 E_3}} & & & \\ \frac{-\nu}{\sqrt{E_1 E_3}} & \frac{-\nu}{\sqrt{E_2 E_3}} & \frac{1}{E_3} & & & \\ & & & \frac{2(1+\nu)}{\sqrt{E_2 E_3}} & & \\ & & & & \frac{2(1+\nu)}{\sqrt{E_1 E_3}} & \\ & & & & & \frac{2(1+\nu)}{\sqrt{E_1 E_2}} \end{bmatrix} \quad (4)$$

Pouya and Reiffsteck (2003) remarked that some of Bohler's (1975) data on the Young's modulus of various soils could be fitted by (4); moreover, they demonstrated that this assumption allows simplifying foundation modelling. The theoretical features of $\Psi$-type materials have been thoroughly explained in Pouya and Zaoui (2006) and Pouya (2007a). For $\Phi_4$-type materials, Pouya (2007b) demonstrated that a closed-form expression for Green's function (displacement solution for a point force within an infinite medium) can be derived, which enables generating an explicit solution for many classical elasticity problems, as well as developing numerical Boundary Element methods for $\Phi$-type materials.

## 3. APPLICATION TO SEDIMENTARY ROCKS

For the study of seismic wave propagation in geological layers, Daley and Hron (1979) developed the concept of an "elliptically anisotropic" medium, as distinct from the ellipsoidal anisotropy considered in the present paper. This concept has been widely used in geophysical studies and Thomsen (1986) undertook an examination within the context of "weak anisotropy" for a large variety of sedimentary rocks. Thomsen (1986) defined four dimensionless parameters $\varepsilon$, $\delta$, $\delta^*$ and $\gamma$, in order to characterise transverse isotropic materials, and provided their values for a wide array of sedimentary rocks. Based on these parameters, the dimensionless elastic coefficients $c_{ij}^*$, defined as $c_{ij}^* = c_{ij}/c_{33}$, can be deduced. These coefficients are given in Table 1 for some of the samples studied by Thomsen.

Rock type and sample depth are presented in the first and second columns of the table, respectively; these two data elements allow identifying the precise corresponding material in the table established by Thomsen. The effort will now be made to compare these materials with the ellipsoidal anisotropy model (2). Since the context here is one of transverse symmetry (Fig. 1), the third condition expressed in (2) is automatically satisfied and the first two conditions become equivalent. The discrepancy between the anisotropy model of these materials and the ellipsoidal model (2) can therefore be measured by the difference between the two sides of the first equality in (2).

Table 1: Dimensionless parameters for several transverse isotropic sedimentary rocks, as deduced from the Thomsen (1986) data (indirect measurements), along with the distance $d$ from the ellipsoidal model. The $\sqrt[4]{c(n)}$ indicator surface applies to a transverse isotropic material with ellipsoidal anisotropy

| Rock | Depth (m) | $c_{11}^*$ | $c_{44}^*$ | $c_{13}^*$ | $c_{12}^*$ | $d$ |
|---|---|---|---|---|---|---|
| Sandstone | 4912.0 | 1.19 | 0.40 | 0.28 | 0.31 | -0.004 |
| | 5481.3 | 1.18 | 0.35 | 0.44 | 0.34 | 0.022 |
| | 6542.6 | 1.16 | 0.34 | 0.32 | 0.36 | -0.037 |
| | 1582.0 | 1.16 | 0.70 | -0.34 | -0.23 | -0.012 |
| Limestone | 5469.5 | 1.11 | 0.34 | 0.32 | 0.34 | -0.027 |
| Mud shale | 7939.5 | 1.16 | 0.33 | 0.45 | 0.43 | 0.019 |
| Clay shale | 5501.0 | 1.67 | 0.27 | 0.99 | 0.49 | 0.094 |
| | 5858.6 | 1.38 | 0.30 | 0.59 | 0.58 | 0.003 |
| | 3511.0 | 1.34 | 0.49 | 0.02 | 0.06 | -0.069 |
| | 450.0 | 1.22 | 0.17 | 0.74 | 0.76 | -0.009 |
| | 650.0 | 1.39 | 0.17 | 0.81 | 0.83 | -0.009 |

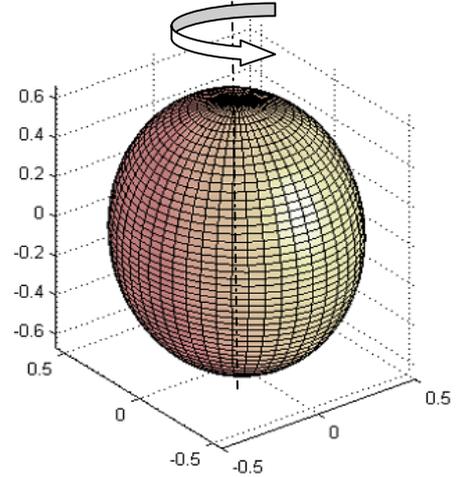

Figure 1 : Ellipsoidal anisotropy with transverse symmetry

Then, by applying $c_{11}=c_{22}$ and $c_{13}=c_{23}$, we are able to define the dimensionless distance $d$ between the transverse isotropic model and the ellipsoidal model as follows:

$$d = \frac{1}{c_{33}} \left( c_{44} - \frac{\sqrt{c_{11} c_{33}} - c_{13}}{2} \right) = c_{44}^* - \frac{\sqrt{c_{11}^*} - c_{13}^*}{2} \quad (5)$$

The value of $d$ calculated for the materials listed in Table 1 is presented in the last column of the table. The parameter $c_{11}^*$ serves to describe the anisotropy. As an example, it may be noticed that for the clay shale sample at a depth of 5,858.6 m, with distinguishable anisotropy, i.e. $c_{11}^* = 1.38$, the assumption of an ellipsoidal model induces an error of just 0.3% ($d = 0.003$). The other rows of the table indicate that despite the pronounced anisotropy, distance to the ellipsoidal model remains relatively small. The mean value of $d$ calculated for all sandstone, limestone, mud shale, clay shale and shale samples (about 25 in all) presented in the Thomsen table (1986) equals approximately 0.03. The ellipsoidal model therefore seems to provide a good fit for the anisotropic parameters of a variety of sedimentary rocks.

## 4. APPLICATION TO FRACTURED ROCK MASSES

Numerical homogenization represents a current method for determining fractured rock mass properties (Pouya and Ghoreychi, 2001; Min and Jing, 2003; Chalhoub, 2006). According to this method, it proves easier to prescribe simple loads, such as simple compression or shear, along different directions and then calculate the corresponding modulus value. Fitting numerical results to an ellipsoidal model simplifies both data analysis and interpretation, in addition to reducing the number of parameters to be determined and providing effective approximate models for certain varieties of rock masses. Moreover, it yields an estimation of elastic parameter values in those directions not accessible through numerical simulation methods.

Chalhoub (2006) offers a study example of a limestone rock mass slope at the border of a main road in Lebanon (see Fig. 2).

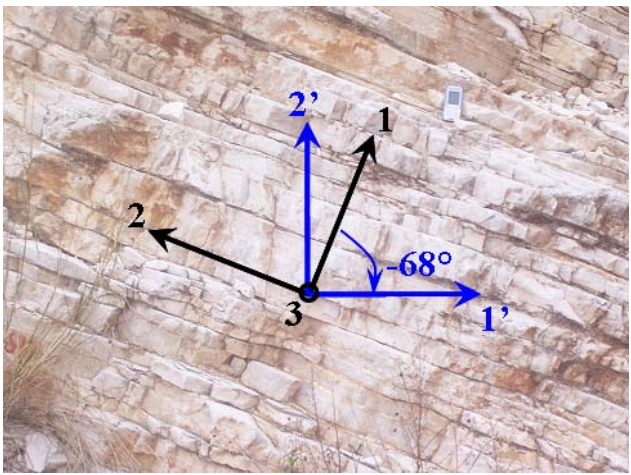

Figure 2: Studied sedimentary rock mass (Chalhoub, 2006)

The rock mass geometry illustrated in this figure reveals the presence of two main sets of fractures, whose geometrical and mechanical properties are listed in Table 2. The geometry indicates that the rock mass exhibits orthotropic behaviour in directions 1, 2 and 3. The parameters obtained for the rock mass by homogenisation are listed in Table 3.

Using the calculated 2D elastic parameters in association with matrix transformation principles, the value of Young's modulus can be calculated in different directions. Figure 3 shows the indicator diagram of both $E$ and $\sqrt[4]{E}$ and allows stating that $\sqrt[4]{E}$ can be well described by an ellipse.

Table 2: Geometrical and mechanical properties of the rock mass

| Joint filling material (limestone) | |
|---|---|
| $l_1$: infinite | $L_2=45$ |
| $d_1=19$ | $d_2=39$ |
| $K_n=2870$, $K_t=768$ | |
| **Rock (clay)** | |
| $E=20000$ | $\nu=0.25$ |

$l(cm)$: average fracture length; $d(cm)$: average fracture spacing; $K_n(MPa/m)$: normal fracture stiffness; $K_t(MPa/m)$: fracture shear stiffness; $E(MPa)$: Young's modulus of the rock; $\nu$: Poisson's ratio of the rock

Table 3: Calculated elastic compliance parameters

| Direction | $E_1(MPa)$ | $E_2(MPa)$ | $E_3(MPa)$ | $G_{12}(MPa)$ | $\nu_{31}$, $\nu_{32}$ | $\nu_{13}$, $\nu_{12}$ | $\nu_{23}$, $\nu_{21}$ |
|---|---|---|---|---|---|---|---|
| 1-2 | 456 | 960 | $2.10^3$ | 94 | 0.25 | 0.034 | 0.06 |

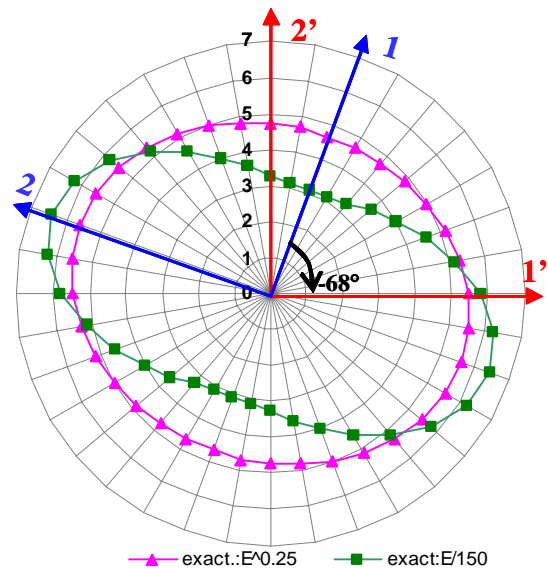

Figure 3: Indicator diagrams of $E$ and $\sqrt[4]{E(n)}$ obtained for a rock mass studied using the numerical homogenization technique (Chalhoub, 2006)

This result is promising and suggests application of the ellipsoidal model to fit numerical data. The 3D model $\Psi$ with four parameters was chosen for this purpose. Parameters $E_1$, $E_2$ and $E_3$, which correspond to the three orthotropic directions, can be deduced directly from Table 3 results. The Poisson's ratio $\nu$ of the model in (4) can then be deduced as an approximation, by means of the following formulae, taken from the values in Table 3:

$$\nu = \sqrt[3]{\sqrt{\nu_{12}\nu_{21}}\sqrt{\nu_{13}\nu_{31}}\sqrt{\nu_{23}\nu_{32}}} \qquad (6)$$

We obtained as a result: $\nu = 0.072$. 2D numerical calculations provide all the elastic compliance parameters, with the exception of elastic compliances $s_{44}$ and $s_{55}$ (Min and Jing, 2003). By using the 3D ellipsoidal model in (4), these parameters can be expressed by a combination of three Young's moduli and one Poisson's ratio $\nu$. This result may be considered a helpful approximation of these parameters, which are inaccessible by means of 2D numerical homogenization methods.

## 5. CONCLUSION

The concept of ellipsoidal anisotropy seems to offer an attractive guideline for the phenomenological modelling of anisotropic elasticity of geomaterials, soils, rocks and rock masses; it simplifies data analysis and serves to create models with a reduced number of parameters and interesting theoretical properties. As discussed in Pouya (2007a), this concept also corresponds well to models of anisotropic elastic tensors obtained for micro-cracked materials using the micro-macro approach. The advantages offered in terms of handling geotechnical problems have already been discussed from a theoretical perspective in Pouya and Reiffsteck (2003) and in Pouya and Zaoui (2006). This paper has shown that such models could also provide a good fit for the experimental or numerical homogenization data on certain varieties of sedimentary rocks. The potential application of these models to wave propagation in rocks, for use in seismic analysis, is one of the promising avenues opened by the present work.

## 6. REFERENCES

Boehler, J.-P. 1975. Contribution théorique et expérimentale à l'étude des milieux plastiques anisotropes. *Thèse d'état Institut de Mécanique de Grenoble*, France.

Chalhoub, M. 2006. Contributions of numerical homogenization methods on the rock mass classifications, (Apports des méthodes d'homogénéisation numérique sur la classification des Massifs Rocheux Fracturés), Ph.D thesis, Ecole des Mines des Paris, 216 p.

Chalhoub, M., Pouya, A.: A geometrical approach to estimate the mechanical Representative Elementary Volume of a fractured rock mass. *First Euro Mediterranean in Advances on Geomaterials and Structures* – Hammamet, Tunisia, 3-5 May 2006.

Daley, P.F. & Hron, F. 1979. Reflection and transmission coefficients for seismic waves in ellipsoidally anisotropic media. Geophysics, 44: 27-38.

Francois M, Geymonat G., Berthaud Y. 1998. Determination of the symmetries of an experimentally determined stiffness tensor : Application to acoustic measurements, *Int. J. Solids Structures*, Vol. 35, Nos. 31-32, pp. 4091-4106.

Homand, F., Morel, E., Henry, J.-P., Cuxac, P. & Hammade, E. 1993. Characterization of the Moduli of Elasticity of an Anisotropic Rock Using Dynamic and Static Methods. *Int. J. Rock Mech. Min. Sci. & Geomech. Abstr.*, Vol. 30, No. 5: 527-535.

Min, K.B. & Jing, L. 2003. Numerical determination of the equivalent elastic compliance tensor for fractured rock masses using the distinct element method. *Int. J. Rock Mech. Min. Sci.*, 40 : 795-816.

Peres Rodrigues, F. & Aires-Barros, L. 1979. Anisotropy of endogenetic rocks- Correlation between micropetrographic index, ultimate strength an modulus of elasticity ellipsoids. *Proc. 2$^{nd}$ Congress of the ISRM*, Belgrade, 1-23.

Pouya, A. 2000. A transformation of the problem of linear elastic structure for application to inclusion problem and to Green functions. *Comptes-Rendus de l'Académie des Sciences de Paris*, t. 328, Série II b, pp. 437-443.

Pouya, A. Ghoreychi, M. 2001. Determination of rock mass strength properties by homogenization., *Int. J. Numer. Anal. Meth. Geomech.*, 25, (2001), 1285-1303.

Pouya, A. & Reiffsteck, Ph. 2003. Analytical solutions for foundations on anisotropic elastic soils. *Proceedings of International Symposium FONDSUP*, Presses des Ponts et Chaussées, Paris.

Pouya, A., Zaoui, A., 2006. A transformation of elastic boundary value problems with application to anisotropic behavior", Int. Journal of Solids and Structures, 43 (2006) 4937-4956.

Pouya A,. 2007a. Ellipsoidal anisotropies in linear elasticity - Extension of Saint Venant's work to phenomenological modelling of materials., *Int. Journal of* Damage Mechanics, *2007, Issue 1, Vol. 16, pp. 95-126.*

Pouya A,. 2007b. Fonction de Green pour les matériaux à anisotropie ellipsoïdale. *Comptes-Rendus Mécanique, Acad. Sciences de Paris, in press.*

Saint Venant, B. (de) 1863. Sur la distribution des élasticités autour de chaque point d'un solide ou d'un milieu de contexture quelconque, particulièrement lorsqu'il est amorphe sans être isotrope. *Journal de Math. Pures et Appliquées*, Tome VIII (2$^{ème}$ série) pp. 257-430.

Thomsen, L. 1986 Weak elastic anisotropy. Geophysics, Vol. 51 No. 10 : 1954-1966.